\documentclass{Interspeech}

\interspeechcameraready

\title{Enhancing Generalization of Speech Large Language Models with Multi-Task Behavior Imitation and Speech-Text Interleaving}

\author[affiliation={1,2}]{Jingran}{Xie}
\author[affiliation={1,2}]{Xiang}{Li}
\author[affiliation={2}]{Hui}{Wang}
\author[affiliation={2}]{Yue}{Yu}
\author[affiliation={2}]{Yang}{Xiang}
\author[affiliation={3}]{Xixin}{Wu}
\author[affiliation={1,3,*}]{Zhiyong}{Wu}

\affiliation{Shenzhen International Graduate School}{Tsinghua University}{China}
\affiliation{}{Pengcheng Laboratory}{China}
\affiliation{}{The Chinese University of Hong Kong}{China}
\email{\{xjr21, lix23\}@mails.tsinghua.edu.cn, \{wangh06, yuy, xiangy\}@pcl.ac.cn, \{wuxx, zywu\}@se.cuhk.edu.hk}

\usepackage{comment}
\usepackage{multirow}
\usepackage{threeparttable}
\usepackage{xcolor}
\definecolor{figure_green}{rgb}{0.0, 0.690, 0.314}
\definecolor{figure_blue}{rgb}{0.125, 0.467, 0.710}

\begin{document}

\maketitle
\renewcommand{\thefootnote}{\fnsymbol{footnote}}
\footnotetext[1]{Corresponding author.}

\begin{abstract}
Large language models (LLMs) have shown remarkable generalization across tasks, leading to increased interest in integrating speech with LLMs. These speech LLMs (SLLMs) typically use supervised fine-tuning to align speech with text-based LLMs. However, the lack of annotated speech data across a wide range of tasks hinders alignment efficiency, resulting in poor generalization. To address these issues, we propose a novel multi-task 'behavior imitation' method with speech-text interleaving, called MTBI, which relies solely on paired speech and transcripts. By ensuring the LLM decoder generates equivalent responses to paired speech and text, we achieve a more generalized SLLM. Interleaving is used to further enhance alignment efficiency. We introduce a simple benchmark to evaluate prompt and task generalization across different models. Experimental results demonstrate that our MTBI outperforms SOTA SLLMs on both prompt and task generalization, while requiring less supervised speech data.
\end{abstract}

\section{Introduction}
\label{intro}
In recent years, large language models (LLMs) \cite{achiam2023gpt, touvron2023llama2} have demonstrated remarkable capabilities, particularly in zero-shot scenarios, showcasing their ability to generalize across a wide range of tasks.
To leverage this ability for speech, substantial progress \cite{zhang2023speechgpt, shu2023llasm, chu2023qwen, chu2024qwen2, xie2024mini} has been made in the development of speech large language models (SLLMs), which process spoken inputs and generate text responses following textual task prompts.
The common approach \cite{shu2023llasm, chu2023qwen, chu2024qwen2} in developing these models involves aligning speech with the text embedding space of LLMs through supervised fine-tuning (SFT) with labeled speech data. This alignment enables the LLM to understand speech content and leverage its generalization ability to handle various tasks based on speech input. Due to the limited scale of speech data, some tricks like multi-task learning \cite{radford2019language, chen2024multi} and multi prompt training \cite{chu2024qwen2} are often used to enhance SLLM's generalization. 
However, despite promising results, some researchers \cite{wang2023blsp} have noted that the SFT-based approach degrades the model’s ability to follow text prompts effectively. Our experiments also indicate that SLLMs tend to exhibit poorer generalization capabilities compared to purely text-based models on zero-shot tasks (Unseen task during SFT). We believe this limitation arises from the scarcity of annotated speech data. Even the most advanced SLLM, such as Qwen2-audio \cite{chu2024qwen2}, is SFT on a limited number tasks, which are far fewer than the wide range encountered in real-world scenarios. 
The inherent difficulty and cost of annotating speech data across diverse tasks make such extensive coverage unlikely.

To address these challenges, we draw inspiration from successful SLLMs \cite{wang2023blsp, chu2024qwen2, das2024speechverse, zeng2024interleave1}, and propose a novel method that leverages multi-task "behavior imitation" training with interleaving, referred to as MTBI. Our motivation is to achieve a high level of alignment between speech and text modalities on easily accessible ASR data, without reliance on large amount of hard-to-obtained supervised data. This well alignment allows the SLLM to fully leverage the generalization ability of the LLM.
We train the SLLM to generate responses equivalent to the text-based LLM when given the same speech and transcription, just like training SLLMs to imitate the behavior of text LLM.
Additionally, we also incorporate multi-task learning method, which performs behavior imitation across diverse task distributions, to further enhance the generalization.
Inspired by recent text-to-speech (TTS) work \cite{zeng2024interleave1, nguyen2024interleave2}, we introduce an interleaving technique that mix speech and text as inputs, effectively improving the alignment between two modalities. 
%Our experimental results demonstrate that this approach improves the alignment efficiency and overall performance of the model in handling both speech and text inputs.

Given the lack of attention to the generalization abilities of SLLMs, we follow the natural language processing (NLP) generalization evaluation methods \cite{mao2023gpteval, zhou2023ifeval, gao2024llm}, to present a simple benchmark to evaluate the generalization capabilities of SLLMs, assessing their adaptability to various prompts and zero-shot tasks. 
% This benchmark offers a clear and systematic way to measure real-world applicability.
In summary, our contributions are as follows:
\begin{itemize}
    \item We propose a novel multi-task behavior imitation training method that leverages easily accessible transcriptions, reducing the reliance on extensive annotated speech data. This method significantly enhances the zero-shot generalization capability of SLLMs, enabling them to perform effectively across a wide range of tasks.
    \item  We introduce an interleaving technique that combines speech and text as inputs. This enhances the alignment efficiency between speech and text modalities, greatly improving the generalization of the SLLM.
    \item We design a benchmark to systematically evaluate SLLMs’ generalization to different prompts and zero-shot tasks, providing a clear framework for comparing model performance and advancing the field. Through experiments, we demonstrate that our MTBI achieves or exceeds the performance of SOTA SLLMs in terms of prompt and task generalization.
\end{itemize}

\begin{figure*}[h]
  \centering
  \includegraphics[width=0.95\linewidth]{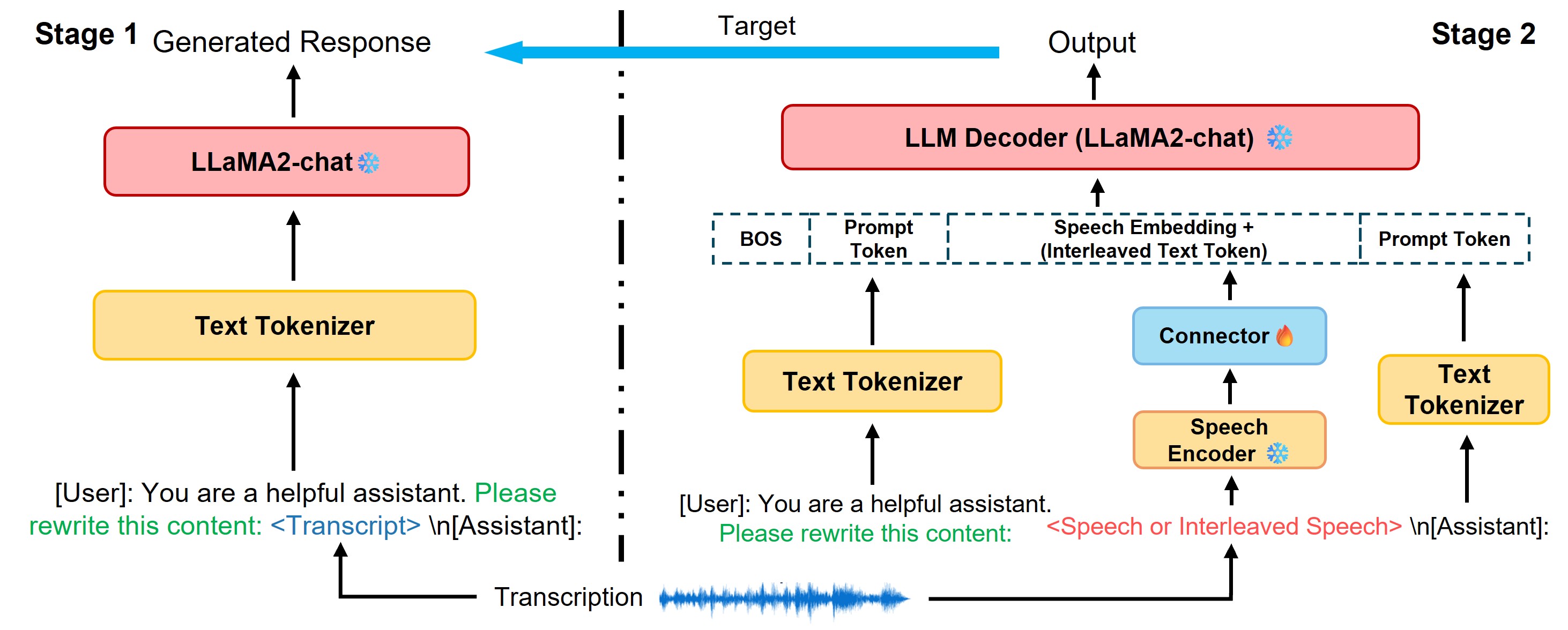}
  \caption{Overview of the SLLM architecture. The training process is conducted in two stages using the same text LLM. In stage 1, we use the LLM to generate responses based on the \textcolor{figure_green}{task prompts } and \textcolor{figure_blue}{transcripts} of speech data. Then, we train the SLLM model with behavior imitation that use the same \textcolor{figure_green}{task prompt} and \textcolor{red}{corresponding speech (or interleaved speech)} to predict the generated response of the first stage. In stage 2, we only train the connector to align the speech features into textual space.}
  \label{fig:model}
  \vspace{-2mm}
\end{figure*}

\section{Methodology}
\label{model}
\subsection{Model Architecture}
We train MTBI on a simple and popular vanilla speech LLM framework \cite{chu2023qwen, chu2024qwen2, xie2025leveraging} as shown in Figure~\ref{fig:model}, right part. This architecture integrates an frozen speech encoder, a trainable modality connector and a frozen text LLM to generate response.  

\textbf{Speech Encoder} 
The speech encoder is responsible for converting raw audio signals into a meaningful representation that contains well-organized linguistic and acoustic features. 
We use the WavLM Large \cite{chen2022wavlm} as the speech encoder, which consists of 24 Transformer layers. We obtain the speech representation from the last layer. To reduce the training costs, we froze the speech encoder throughout the entire training.

\textbf{Connector} 
We introduce a small trainable connector \cite{zhang2023llama, li2023blip} to link the speech and text modalities, which can efficiently map speech representation to the textual space of LLM with minimal computational cost.
Since speech embeddings are typically much longer than text embeddings, we use a CNN based subsampler connector to reduce the length discrepancy between two modalities. 
This connector functions as an intermediary between the speech encoder and the LLM decoder, efficiently fusing speech embeddings with textual input and facilitating smooth interactions between the two modalities.

\textbf{LLM Decoder} 
To generate text responses, we employ LLaMA2-7B-chat \cite{touvron2023llama2} as the LLM decoder, which is chosen for its strong conversational abilities. The decoder receives text task prompts and processed speech embeddings from the encoder and connector, then generates a coherent output.
To prevent catastrophic forgetting of internal knowledge, LLM decoder is frozen throughout the training process.

\subsection{Behavior Imitation}
\label{bi}
\subsubsection{Behavior Imitation Training}
As discussed in the introduction, our objective is let the LLM decoder produce consistent responses when presented on speech and its corresponding transcription. Achieving this consistency would indicate that the speech is treated just like the text which also means the speech is well-aligned with the LLM's text embedding space. 
Therefore, we propose behavior imitation for this alignment, which consists of two stages: first, the LLM decoder generates a text response based on task prompts and transcriptions; then, we use the same prompts and paired speech to train the SLLM to predict the generated text.
The method offers two key advantages:
1) By training the model to imitate the behavior of a text LLM, we encourages the model to process speech content similarly to text. This training enhances the alignment between speech and LLM's textual space, allowing the SLLM to leverage the LLM’s knowledge, improving generalization.
2) Our approach relies solely on transcriptions, which are more cost-effective to obtain compared to other annotations, such as intent, emotions, etc. 
Meanwhile, by using easily accessible text transcriptions, our MBTI can be applied to any speech data at a low cost, enhancing model generalization.

\subsubsection{Multi-Task Learning}
To further improve generalization, we use multi-task learning, training the model across various tasks, similar to those used in other SLLMs \cite{shu2023llasm, chu2023qwen, chu2024qwen2}.  This training approach enables the model to learn shared representations generalizable across different domains and contexts. 
% The input format for our behavior imitation is like this:
% \textit{[User]: You are a helpful assistant. $<$Task prompt$>$: $<$transcript/speech$>$\textbackslash n$[$Assistant$]$:}
Our focus is on content-related tasks that are also used for SLLMs' SFT, like intent classification (IC), slot filling (SF), and speech translation (ST), aiming to improve speech and text alignment.
We construct system and task prompts as illustrated in Figure~\ref{fig:model}. 
To demonstrate the generalization capability of our method, we restrict training to a single task prompt per task, omitting the strategy of generating similar prompts to enhance generalization.

\subsubsection{Constructed Tasks}
\label{constructed}
Although behavior imitation can theoretically perform on any speech dataset as long as the transcript is available, we observe that the quality of speech data and accuracy of transcription play a critical role in the model’s performance. 
To further investigate this, we try to replace the lower-quality dataset with the higher-quality supervised dataset on the same task.
The results show significant improvements in model performance. Building on this observation, we choose high-quality ASR datasets like LibriSpeech for training. Since LibriSpeech mostly consists of audiobooks, it is less suited for tasks like IC, or SF that we previously explored. Therefore, we decided to construct several new content-related tasks:

\textbf{Continuation} 
The continuation task encourages the model to generate a continuation of the speech or transcription, enhancing its ability to understand the input content and produce consistent output across both modalities.

\textbf{Rewriting} 
The rewriting task prompts the model to rephrase the input content, which aids in improving its ability to understand language nuances.

\textbf{Selecting}
% The selecting task challenges the model to identify verbs and nouns from a sentences. 
This task asks the model to select all verbs or nouns in the input utterance.
Since verbs and nouns are usually key components of a sentence, carrying the most significant meaning. This task trains the model to identify these key components, enhancing its ability to focus on important content.

\subsubsection{Task Setup}
\label{task-setup} 
Recognizing the importance of accurate transcription for speech-related tasks, we retain the ASR SFT task to ensure the model's transcription capabilities. The final multi-task setup includes ASR SFT and three new constructed tasks. For a fair comparison with multi-task SFT, we also train behavior imitation on IC, SF, and ST supervised dataset alongside ASR SFT.

\subsection{Speech-Text Interleaving}
Inspired by current text-to-speech (TTS) methods \cite{zeng2024interleave1, nguyen2024interleave2}, we incorporate an interleaving strategy to encourage the model to learn an alignment between speech and text, facilitating the transfer of text-based knowledge to speech modality.

To avoid reliance on the time-aligned speech-text parallel datasets, we achieve interleaving over the transcription of speech data with the TTS method. We randomly select a contiguous segment corresponding to approximately 40–60\% of the input transcription’s length and replace that portion with the corresponding speech segment. To avoid confusing the model, interleaving is not applied to the text prompt, and only a single speech segment is interleaved per input. For example, a transcription might be transformed into “Hope you $<$Interleaved Speech$>$.” 
We employ Cosyvoice 2 \cite{du2024cosyvoice2} to synthesize the speech segments from selected text spans. Subsequently, the text and speech segments are encoded separately using the LLM tokenizer and a dedicated speech encoder.
During each training step, we determine whether to apply the interleaving strategy to the batch based on a predefined probability for all tasks.

\section{Test Set}
\label{test-set}
Current research on generalization for SLLMs is relatively scarce. To address this gap, we draw inspiration from established evaluation methods in NLP \cite{zhou2023ifeval, gao2024llm}. We construct a preliminary test set to assess model's generalization across two dimensions: prompt generalization and task generalization.

\subsection{Prompt Generalization}
We assess the model's prompt generalization using a diverse set of prompts on a common task, such as ASR. Specifically, we randomly select 10 samples with transcripts from diverse datasets. GPT-4o \cite{openai2024gpt4o} is used to generate a wide variety of prompt descriptions to assess the prompt generalization of each model. The total number of clips is 10, and the total number of prompts is 100. The model's response to these prompts is then evaluated by GPT-4o, which checks whether the generated output follows the prompt correctly and calculates the accuracy.

\subsection{Task Generalization}
A key characteristic of LLMs is their zero-shot learning ability, also known as emergent ability, to perform tasks not encountered during training. We define this capability as task generalization. To validate this property, we design two tasks: speaker role inference and mathematical reasoning, which are typically not included for most of SLLMs during the SFT.

\textbf{Speaker role}
In the speaker role inference task, we assess the model's ability to infer the speaker's role from the speech content. We leverage GPT-4o to generate 60 diverse utterances associated with five roles (student, teacher, doctor, police, engineer), which are then synthesized into speech using CosyVoice 2. Models were prompted to select one role from five options based on the content of the speech. This task evaluates both the instruction following and reasoning abilities of SLLMs. The performance is measured by inference accuracy.

\textbf{GSM8K}
For the mathematical reasoning task, we assess the model’s ability to solve arithmetic and reasoning problems presented in spoken form. Since the backend LLM are not specific trained for mathematical problems, we construct the dataset based on GSM8K \cite{cobbe2021training}, which contains primary school-level math problems. To ensure compatibility for TTS synthesis, we filter out problems containing complex mathematical formulas, symbols, or notations that would be challenging for TTS models to generate naturally. Following this preprocessing step, we select 1,100 well-structured problems and synthesize them into speech using CosyVoice 2. In line with standard LLM math evaluation practices \cite{hendrycks2021measuring}, model performance is evaluated solely on the accuracy of the final numerical result.

\begin{table*}[h]
  \caption{The main results of our experiments are presented by comparing the performance of cascaded models, popular SLLMs, and several configurations of our vanilla baseline model. The evaluation tasks are traditional ASR and our generalization test set.}
  \label{tab:main}
  \centering
  \begin{threeparttable}
  \begin{tabular}{l|cc|cc|c}
    \toprule
    & \textbf{ASR WER} $\downarrow$ & \textbf{Prompt Generalization} $\uparrow$ & \textbf{GSM8K} $\uparrow$ & \textbf{GSM8K 1-shot} $\uparrow$ & \textbf{Speaker Role} $\uparrow$ \\
    \midrule
    \multicolumn{6}{c}{Cascaded Models} \\
    \midrule
    LLaMA2 + GT      & -  &  -  & 36.5   & 59.6 & 88.7\\
    LLaMA2 + ASR     & -  &  -  & 21.4   & 40.5 & 73.3\\
    \midrule
    \multicolumn{6}{c}{Speech Large Language Model} \\
    \midrule
    ASR SFT       & 1.97  & 97*    & 0.0   & -  & 3.67\\
    Multi-task SFT & 3.12  & 53     & 0.0   & -  & 13.7\\
    Ours w/ ASR$^1$, IC, SF, ST    & 6.97  & 78     & 9.1  & 11.5 & 65.3\\
    Ours w/ Constructed Tasks   & 3.97  & \textbf{85}     & \textbf{20.1}  & \textbf{25.5} & \textbf{75.3}\\
    \midrule
    % \multicolumn{6}{c}{Speech Large Language Model} \\
    % \midrule
    BLSP \cite{wang2023blsp}             & 22.4  & 70       & 1.17  & 1.53 & 50.7\\
    Qwen-audio-chat \cite{chu2023qwen}  & 7.1  &  59       & 0.24  & -    & 27.0\\
    Qwen2-audio-instruct \cite{chu2024qwen2}  & \textbf{2.4}  & 83   & 10.5  & 6.1  & 55.7 \\
    \bottomrule
  \end{tabular}
\begin{tablenotes}
    \footnotesize
    \item [*] The ASR SFT tends to overfit to ASR tasks, causing the model to prioritize transcription over the prompt, resulting in artificially high accuracy on our ASR-based prompt generalization evaluation.
    \item [1] As mentioned in Section~\ref{task-setup}, the ASR task in our proposed method is a SFT not a behavior imitation training.
\end{tablenotes}
\end{threeparttable}
\vspace{-2mm}
\end{table*}

\section{Experiment}
\subsection{Experimental Setup}
\label{setup}
For multi-task learning, we use several open-source supervised datasets: LibriSpeech \cite{panayotov2015librispeech} for ASR, Audio Snips \cite{coucke2018snips} for IC, Fluent Speech Commands \cite{lugosch2019fluent} for SF, and CoVoST 2 \cite{ardila2019commonvoice, wang2021covost} for ST. We balance these datasets to ensure a similar proportion across tasks. To maintain a fair comparison, our MTBI uses the same dataset and ratio, with ASR using SFT and others using behavior imitation. For three constructed tasks, we use only LibriSpeech 960 hours data and set the same ratio for all tasks with an additional ASR SFT task. The interleaving probability is 40\% for each batch. Training uses a learning rate of 1e-5, with a batch size of 16 on 4 NVIDIA A100 80G GPUs. The training process runs for one epoch.

For evaluation, we use the standard ASR evaluation on the LibriSpeech test-clean dataset with word error rate (WER) as the evaluation metric. Details of the generalization evaluation are in Section~\ref{test-set}. During inference, LLM applies a temperature of 0.7 and top-p sampling with a 0.85 threshold, setting a maximum token length of 100 (200 for math tasks) to minimize hallucinations.

\subsection{Comparison with Cascaded Models}
\label{alignment}
To evaluate the upper-bound performance achievable by a text-based Large Language Model (LLM), we utilize LLaMA2 with ground truth (GT) transcriptions. We then compare this to a cascaded system consisting of Whisper Large V3 \cite{radford2022whisper}, a SOTA ASR model, followed by LLaMA2 (denoted as LLaMA2 + ASR). 
As shown in Table~\ref{tab:main}, our proposed method achieves comparable performance to LLaMA2 + ASR on the speaker role inference task, with a slight performance decrease on the mathematical reasoning task. 
Through case studies, we find that math questions require high accuracy in the transcribed results, as even a small error in transcription can lead to incorrect final answers. This explains the relatively lower performance of both the LLM + ASR system and our method compared to LLaMA2 + GT. Nevertheless, the results demonstrate that our method effectively leverages the inter-knowledge within LLMs, showcasing promising emergent abilities. Additionally, as a text-based LLM, our approach exhibits the 1-shot capability, which enhances multi-modal reasoning abilities with text samples.
This finding suggests significant potential for further improvement.

\subsection{Comparison with SOTA SLLM}
Due to differences in model architecture, training setups, and datasets, the comparison with SLLMs primarily aims to investigate the performance gap between our method and the current SOTA SLLMs. As shown in Table~\ref{tab:main},
our method achieves comparable prompt generalization and outperforms Qwen2-audio in zero-shot tasks, using only 960 hours of ASR data and a single task prompt, much less than what is typically used by mainstream SLLMs. This highlights the strong generalization capability of our MTBI, emphasizing its potential.

\subsection{Comparison on Baseline Model}
We compare the performance of the single ASR SFT model, multi-task SFT model, and our multi-task behavior imitation model. As shown in Table~\ref{tab:main}, while the ASR and multi-task SFT models perform well on ASR WER, they exhibit limited generalization across prompts, especially on math tasks, where there are no options in the prompts to randomly guess. In contrast, our MBTI model, trained with the same dataset as the multi-task SFT performs much better on math and role inference tasks, demonstrating its improved generalization. As discussed in Section~\ref{constructed}, supervised data quality may impact model performance. Therefore, we train MTBI on high quality ASR data LibriSpeech with constructed tasks. The results show that the performance is significantly enhanced, supporting our observation and proving that our method can significantly improve the generalization of SLLM.

\begin{table}\scriptsize 
  \caption{Ablation study. "C," "R," "S," and "I" correspond to continuation, rewriting, selecting tasks, and interleaving.}
  \label{tab:ablation}
  \centering
  \begin{tabular}{ccccccc}
    \toprule
     \textbf{ASR SFT} & \textbf{C} & \textbf{R} & \textbf{S} & \textbf{I}  & \textbf{Prompt Generalization} $\uparrow$ & \textbf{WER} $\downarrow$\\
    \midrule
    % \checkmark  & \times      & \times      & \times        & \times    & -  & 1.97\\
    $\times$  & $\checkmark$  & $\times$      & $\times$        & $\times$    & 46 & 17.1\\
    $\times$  & $\times$      & $\checkmark$  & $\times$        & $\times$     & 35 & 12.5\\
    $\times$   & $\times$       & $\times$       & \checkmark    & $\times$     & 36 & 15.4\\
    $\times$       & \checkmark  & \checkmark  & \checkmark    & $\times$     & 60 & 9.70\\
    \checkmark  & \checkmark  & \checkmark  & \checkmark    & $\times$     & 77 & \textbf{3.79} \\
    \checkmark  & \checkmark  & \checkmark  & \checkmark    & \checkmark & \textbf{85} & 3.97\\
    \bottomrule
  \end{tabular}
  \vspace{-2mm}
\end{table}

\subsection{Ablation Study}
\label{ablation}
We perform an ablation study on constructed tasks to evaluate contribution of ASR SFT, multi-task learning, and speech-text interleaving.
Our experiments show that training each task individually results suboptimal performance. However, when training three tasks together with multi-task learning, the model's generalization to prompts improves significantly, underscoring the importance of this approach.
To evaluate the effect of the ASR SFT task, we compare models trained with and without it. The results show that incorporating the ASR SFT task into behavior imitation training significantly reduces WER, allowing the model to capture spoken content more accurately.
Finally, we investigate the role of speech-text interleaving. Removing interleaving leads to a noticeable decrease in the generalization of the model. We believe that interleaving helps the model distinguish boundaries between prompts and inputs based on content, rather than relying on distributional differences between the two modalities. This enables SLLM to integrate speech and text more effectively, improving its generalization.

\section{Conclusion}
In this work, we propose a novel approach to enhance the generalization capabilities of SLLMs through multi-task behavior imitation with speech-text interleaving. Our experiments demonstrate that our method, trained solely on ASR-supervised data using three constructed tasks, significantly boosts the generalization on both prompts and tasks. We also introduce a benchmark to evaluate SLLMs generalization, offering a clearer understanding of their capabilities. Our approach outperforms existing SOTA SLLMs, offering a promising direction for developing more generalized SLLMs. Furthermore, this method lays a foundation for incorporating nonlinguistic speech features, opening the door to the creation of a comprehensive, highly generalized SLLM.

\bibliographystyle{IEEEtran}
\bibliography{mybib}

\begin{thebibliography}{10}
\providecommand{\url}[1]{#1}
\csname url@samestyle\endcsname
\providecommand{\newblock}{\relax}
\providecommand{\bibinfo}[2]{#2}
\providecommand{\BIBentrySTDinterwordspacing}{\spaceskip=0pt\relax}
\providecommand{\BIBentryALTinterwordstretchfactor}{4}
\providecommand{\BIBentryALTinterwordspacing}{\spaceskip=\fontdimen2\font plus
\BIBentryALTinterwordstretchfactor\fontdimen3\font minus \fontdimen4\font\relax}
\providecommand{\BIBforeignlanguage}[2]{{%
\expandafter\ifx\csname l@#1\endcsname\relax
\typeout{** WARNING: IEEEtran.bst: No hyphenation pattern has been}%
\typeout{** loaded for the language `#1'. Using the pattern for}%
\typeout{** the default language instead.}%
\else
\language=\csname l@#1\endcsname
\fi
#2}}
\providecommand{\BIBdecl}{\relax}
\BIBdecl

\bibitem{achiam2023gpt}
J.~Achiam, S.~Adler, S.~Agarwal, L.~Ahmad, I.~Akkaya, F.~L. Aleman, D.~Almeida, J.~Altenschmidt, S.~Altman, S.~Anadkat \emph{et~al.}, ``Gpt-4 technical report,'' \emph{arXiv preprint arXiv:2303.08774}, 2023.

\bibitem{touvron2023llama2}
H.~Touvron, L.~Martin, K.~Stone, P.~Albert, A.~Almahairi, Y.~Babaei, N.~Bashlykov, S.~Batra, P.~Bhargava, S.~Bhosale \emph{et~al.}, ``Llama 2: Open foundation and fine-tuned chat models,'' \emph{arXiv preprint arXiv:2307.09288}, 2023.

\bibitem{zhang2023speechgpt}
D.~Zhang, S.~Li, X.~Zhang, J.~Zhan, P.~Wang, Y.~Zhou, and X.~Qiu, ``Speechgpt: Empowering large language models with intrinsic cross-modal conversational abilities,'' \emph{arXiv preprint arXiv:2305.11000}, 2023.

\bibitem{shu2023llasm}
Y.~Shu, S.~Dong, G.~Chen, W.~Huang, R.~Zhang, D.~Shi, Q.~Xiang, and Y.~Shi, ``Llasm: Large language and speech model,'' \emph{arXiv preprint arXiv:2308.15930}, 2023.

\bibitem{chu2023qwen}
Y.~Chu, J.~Xu, X.~Zhou, Q.~Yang, S.~Zhang, Z.~Yan, C.~Zhou, and J.~Zhou, ``Qwen-audio: Advancing universal audio understanding via unified large-scale audio-language models,'' \emph{arXiv preprint arXiv:2311.07919}, 2023.

\bibitem{chu2024qwen2}
Y.~Chu, J.~Xu, Q.~Yang, H.~Wei, X.~Wei, Z.~Guo, Y.~Leng, Y.~Lv, J.~He, J.~Lin \emph{et~al.}, ``Qwen2-audio technical report,'' \emph{arXiv preprint arXiv:2407.10759}, 2024.

\bibitem{xie2024mini}
Z.~Xie and C.~Wu, ``Mini-omni: Language models can hear, talk while thinking in streaming,'' \emph{arXiv preprint arXiv:2408.16725}, 2024.

\bibitem{radford2019language}
A.~Radford, J.~Wu, R.~Child, D.~Luan, D.~Amodei, I.~Sutskever \emph{et~al.}, ``Language models are unsupervised multitask learners,'' \emph{OpenAI blog}, vol.~1, no.~8, p.~9, 2019.

\bibitem{chen2024multi}
S.~Chen, Y.~Zhang, and Q.~Yang, ``Multi-task learning in natural language processing: An overview,'' \emph{ACM Computing Surveys}, vol.~56, no.~12, pp. 1--32, 2024.

\bibitem{wang2023blsp}
C.~Wang, M.~Liao, Z.~Huang, J.~Lu, J.~Wu, Y.~Liu, C.~Zong, and J.~Zhang, ``Blsp: Bootstrapping language-speech pre-training via behavior alignment of continuation writing,'' \emph{arXiv preprint arXiv:2309.00916}, 2023.

\bibitem{das2024speechverse}
N.~Das, S.~Dingliwal, S.~Ronanki, R.~Paturi, Z.~Huang, P.~Mathur, J.~Yuan, D.~Bekal, X.~Niu, S.~M. Jayanthi \emph{et~al.}, ``Speechverse: A large-scale generalizable audio language model,'' \emph{arXiv preprint arXiv:2405.08295}, 2024.

\bibitem{zeng2024interleave1}
A.~Zeng, Z.~Du, M.~Liu, L.~Zhang, S.~Jiang, Y.~Dong, and J.~Tang, ``Scaling speech-text pre-training with synthetic interleaved data,'' \emph{arXiv preprint arXiv:2411.17607}, 2024.

\bibitem{nguyen2024interleave2}
T.~A. Nguyen, B.~Muller, B.~Yu, M.~R. Costa-Jussa, M.~Elbayad, S.~Popuri, P.-A. Duquenne, R.~Algayres, R.~Mavlyutov, I.~Gat \emph{et~al.}, ``Spirit-lm: Interleaved spoken and written language model,'' \emph{arXiv preprint arXiv:2402.05755}, 2024.

\bibitem{mao2023gpteval}
R.~Mao, G.~Chen, X.~Zhang, F.~Guerin, and E.~Cambria, ``Gpteval: A survey on assessments of chatgpt and gpt-4,'' \emph{arXiv preprint arXiv:2308.12488}, 2023.

\bibitem{zhou2023ifeval}
J.~Zhou, T.~Lu, S.~Mishra, S.~Brahma, S.~Basu, Y.~Luan, D.~Zhou, and L.~Hou, ``Instruction-following evaluation for large language models,'' \emph{arXiv preprint arXiv:2311.07911}, 2023.

\bibitem{gao2024llm}
M.~Gao, X.~Hu, J.~Ruan, X.~Pu, and X.~Wan, ``Llm-based nlg evaluation: Current status and challenges,'' \emph{arXiv preprint arXiv:2402.01383}, 2024.

\bibitem{xie2025leveraging}
J.~Xie, S.~Lei, Y.~Yu, Y.~Xiang, H.~Wang, X.~Wu, and Z.~Wu, ``Leveraging chain of thought towards empathetic spoken dialogue without corresponding question-answering data,'' \emph{arXiv preprint arXiv:2501.10937}, 2025.

\bibitem{chen2022wavlm}
S.~Chen, C.~Wang, Z.~Chen, Y.~Wu, S.~Liu, Z.~Chen, J.~Li, N.~Kanda, T.~Yoshioka, X.~Xiao \emph{et~al.}, ``Wavlm: Large-scale self-supervised pre-training for full stack speech processing,'' \emph{IEEE Journal of Selected Topics in Signal Processing}, vol.~16, no.~6, pp. 1505--1518, 2022.

\bibitem{zhang2023llama}
R.~Zhang, J.~Han, C.~Liu, P.~Gao, A.~Zhou, X.~Hu, S.~Yan, P.~Lu, H.~Li, and Y.~Qiao, ``Llama-adapter: Efficient fine-tuning of language models with zero-init attention,'' \emph{arXiv preprint arXiv:2303.16199}, 2023.

\bibitem{li2023blip}
J.~Li, D.~Li, S.~Savarese, and S.~Hoi, ``Blip-2: Bootstrapping language-image pre-training with frozen image encoders and large language models,'' in \emph{International conference on machine learning}.\hskip 1em plus 0.5em minus 0.4em\relax PMLR, 2023, pp. 19\,730--19\,742.

\bibitem{du2024cosyvoice2}
Z.~Du, Y.~Wang, Q.~Chen, X.~Shi, X.~Lv, T.~Zhao, Z.~Gao, Y.~Yang, C.~Gao, H.~Wang \emph{et~al.}, ``Cosyvoice 2: Scalable streaming speech synthesis with large language models,'' \emph{arXiv preprint arXiv:2412.10117}, 2024.

\bibitem{openai2024gpt4o}
\BIBentryALTinterwordspacing
OpenAI, ``Hello gpt-4o,'' OpenAI, Tech. Rep., 2024. [Online]. Available: \url{https://openai.com/index/hello-gpt-4o}
\BIBentrySTDinterwordspacing

\bibitem{cobbe2021training}
K.~Cobbe, V.~Kosaraju, M.~Bavarian, M.~Chen, H.~Jun, L.~Kaiser, M.~Plappert, J.~Tworek, J.~Hilton, R.~Nakano \emph{et~al.}, ``Training verifiers to solve math word problems,'' \emph{arXiv preprint arXiv:2110.14168}, 2021.

\bibitem{hendrycks2021measuring}
D.~Hendrycks, C.~Burns, S.~Kadavath, A.~Arora, S.~Basart, E.~Tang, D.~Song, and J.~Steinhardt, ``Measuring mathematical problem solving with the math dataset,'' \emph{arXiv preprint arXiv:2103.03874}, 2021.

\bibitem{panayotov2015librispeech}
V.~Panayotov, G.~Chen, D.~Povey, and S.~Khudanpur, ``Librispeech: an asr corpus based on public domain audio books,'' in \emph{2015 IEEE international conference on acoustics, speech and signal processing (ICASSP)}.\hskip 1em plus 0.5em minus 0.4em\relax IEEE, 2015, pp. 5206--5210.

\bibitem{coucke2018snips}
A.~Coucke, A.~Saade, A.~Ball, T.~Bluche, A.~Caulier, D.~Leroy, C.~Doumouro, T.~Gisselbrecht, F.~Caltagirone, T.~Lavril \emph{et~al.}, ``Snips voice platform: an embedded spoken language understanding system for private-by-design voice interfaces,'' \emph{arXiv preprint arXiv:1805.10190}, 2018.

\bibitem{lugosch2019fluent}
L.~Lugosch, M.~Ravanelli, P.~Ignoto, V.~S. Tomar, and Y.~Bengio, ``Speech model pre-training for end-to-end spoken language understanding,'' \emph{arXiv preprint arXiv:1904.03670}, 2019.

\bibitem{ardila2019commonvoice}
R.~Ardila, M.~Branson, K.~Davis, M.~Henretty, M.~Kohler, J.~Meyer, R.~Morais, L.~Saunders, F.~M. Tyers, and G.~Weber, ``Common voice: A massively-multilingual speech corpus,'' \emph{arXiv preprint arXiv:1912.06670}, 2019.

\bibitem{wang2021covost}
C.~Wang, A.~Wu, J.~Gu, and J.~Pino, ``Covost 2 and massively multilingual speech translation.'' in \emph{Interspeech}, 2021, pp. 2247--2251.

\bibitem{radford2022whisper}
\BIBentryALTinterwordspacing
A.~Radford, J.~W. Kim, T.~Xu, G.~Brockman, C.~McLeavey, and I.~Sutskever, ``Robust speech recognition via large-scale weak supervision,'' 2022. [Online]. Available: \url{https://arxiv.org/abs/2212.04356}
\BIBentrySTDinterwordspacing

\end{thebibliography}

\end{document}